\documentclass[10pt,journal]{IEEEtran}
\def\BibTeX{{\rm B\kern-.05em{\sc i\kern-.025em b}\kern-.08em
    T\kern-.1667em\lower.7ex\hbox{E}\kern-.125emX}}
\usepackage{amsmath}
\usepackage{amsthm}
\usepackage{amsfonts}
\usepackage{amssymb}
\usepackage{graphicx}
\usepackage{cite}
\usepackage{subfigure}
\usepackage{balance}
\usepackage{algorithm}
\usepackage{algorithmic}
\usepackage{enumerate}
\usepackage{color}
\usepackage{array}
\usepackage{indentfirst}
\usepackage{multirow}
\usepackage{booktabs}
\usepackage{color}
\usepackage{slashbox}
\usepackage{diagbox}
\usepackage{threeparttable}
\usepackage[svgnames,table]{xcolor}
\usepackage{bm}
\usepackage{subfigure}
\usepackage{epstopdf}

\makeatletter

\newcommand{\Rmnum}[1]{\expandafter\@slowromancap\romannumeral #1@}
\makeatother

\begin{document}
\title{ Deep Residual Network Empowered Channel Estimation for IRS-Assisted Multi-User Communication Systems \vspace{0.2cm} }

\author{ Chang Liu, Xuemeng Liu, Derrick Wing Kwan Ng, and Jinhong Yuan \\
School of Electrical Engineering and Telecommunications, the University of New South Wales, Australia \\
Email: \{chang.liu19, w.k.ng, j.yuan\}@unsw.edu.au, xuemeng.liu.ac@gmail.com \\ 

\thanks{The journal version of this work has been presented in \cite{liu2020deepresidual}. }}


\maketitle

\begin{abstract}
Channel estimation is of great importance in realizing practical intelligent reflecting surface-assisted multi-user communication (IRS-MC) systems. However, different from traditional communication systems, an IRS-MC system generally involves a cascaded channel with a sophisticated statistical distribution, which hinders the implementations of the Bayesian estimators.
To further improve the channel estimation performance, in this paper, we model the channel estimation as a denoising problem and adopt a data-driven approach to realize the channel estimation.
Specifically, we propose a convolutional neural network (CNN)-based deep residual network (CDRN) to implicitly learn the residual noise for recovering the channel coefficients from the noisy pilot-based observations.
In the proposed CDRN, a CNN denoising block equipped with an element-wise subtraction structure is designed to exploit both the spatial features of the noisy channel matrices and the additive nature of the noise simultaneously, which further improves the estimation accuracy.
Simulation results demonstrate that the proposed method can almost achieve the same estimation accuracy as that of the optimal minimum mean square error (MMSE) estimator requiring the knowledge of the channel distribution.
\end{abstract}

\section{Introduction\label{sect: intr}}
Recently, intelligent reflecting surface (IRS) \cite{gong2020towards} has been proposed as one of the key enable techniques for the beyond fifth generation (5G) networks \cite{wong2017key} due to its powerful capability in shaping the wireless channels between the users and the base station (BS) to enhance the system performance.
Generally, an IRS is composed of a large number of passive reflecting elements such that it can alter the reflection of the incident signals through adapting the phase shift of each reflecting element.
In particular, via jointly adjusting the phase shifts of all the passive elements to form a desirable reflection pattern, the IRS can establish a favourable wireless channel to improve the communication quality with a low power consumption \cite{wu2020towards}.
Therefore, the deployments of an IRS in different types of communication systems have been studied to enhance the energy efficiency \cite{cai2020resource} or the spectral efficiency \cite{zhiqiang2020sum, li2020ergodic} of communication networks through passive beamforming techniques, i.e., designing an efficient configuration of phase shifts at the IRS to improve the received signal-to-noise ratio (SNR) at the desired receivers \cite{wu2019intelligent}.

However, the promised performance gains brought by an IRS relies on accurate channel state information (CSI), which is usually not available \cite{li2020performance, li2020hybrid}.
In fact, an indispensable task for realizing IRS-assisted communication systems is channel estimation.
Different from the channel estimation in traditional systems, IRS is generally a passive device which cannot perform training sequence transmission/reception or signal processing. In other words,  the separate channel of IRS-to-user/BS is not available and only a cascaded channel of user-to-IRS-to-BS can be estimated.
More importantly, the inherent cascaded channel brings two main challenges to IRS-assisted communication systems which are listed as follows:
(i) limited channel estimation performance:
Note that the cascaded user-to-IRS-to-BS channel does not follow the conventional Rayleigh fading model. In this case, the optimal minimum mean square error (MMSE) estimator involves a multidimensional integration which is overly computationally intensive for practical implementation.
Meanwhile, the performance of the available linear MMSE (LMMSE) and least square (LS) estimators still has a large performance gap compared with that of the optimal MMSE estimator.
Thus, the channel estimation performance is unsatisfactory for practical IRS-assisted communication systems.
(ii) large channel estimation training overhead:
The IRS generally consists of a large number of elements. Thus, the cascaded channel is with a high dimension and the corresponding channel estimation via conventional methods, e.g., the LS method and the LMMSE method, is computationally costly.
\cite{}
To overcome these challenges, a variety of effective algorithms and schemes have been proposed recently.
For example, a binary reflection controlled LS channel estimation scheme was developed in \cite{mishra2019channel}, where the IRS only switches on one reflecting element and switches off the remaining reflecting elements at each time slot.
Although the BS does not receive any interference from the other reflection elements, it can only obtain a small received SNR for channel estimation.
Besides, to further improve the received SNR, the authors in \cite{jensen2020optimal} proposed to switch on all the reflecting elements of the IRS at each time slot and developed a discrete Fourier transform (DFT) training sequence-based minimum variance unbiased estimation scheme, which achieves satisfactory estimation accuracy.
On the other hand, some initial attempts have been devoted to the design of efficient schemes to reduce the required training overhead for channel estimation.
For example, \cite{chen2019channel} proposed the sparse matrix-based estimation algorithm to offload the exceeding long training overhead.
Moreover, \cite{wang2020channel} developed a three-phase estimation framework to shorten the required time duration for channel estimation by exploiting the redundancy of the reflecting channels.
In addition, deep learning (DL)-based channel estimation schemes have proved their effectiveness in IRS-assisted communication systems. For example, \cite{elbir2020deep} and \cite{liu2020deepdenoising} both adopted convolutional neural network (CNN)-based architectures to address the channel estimation problem for IRS-assisted communication systems. However, these schemes require either sequentially switching on the individual IRS element one-by-one or the deployment of a hybrid passive/active IRS, which brings additional training overheads or hardware costs.
Thus, an effective and practical algorithm which can further improve the estimation accuracy is expected.

Note that the channel estimation problem is essentially a denoising problem \cite{lxm2020deepresidual}. Meanwhile, the DL techniques \cite{xie2019activity, yuan2020learning, liu2020location}, especially the deep residual network (DRN) has powerful capability in denoising for various application scenarios \cite{zhang2017beyond}.
Motivated by this, in this paper, we focus on IRS-assisted multi-user communication (IRS-MC) systems and adopt a DRN to intelligently exploit the channel features to recover the channel coefficients.
Different from the existing DL-based channel estimation schemes \cite{elbir2020deep, liu2020deepdenoising}, this work further improves the channel estimation accuracy without the requirements of additional system deployment.
The main contributions of this work are listed as follows:
\begin{itemize}
\item In contrast to existing channel estimation methods \cite{mishra2019channel, jensen2020optimal, chen2019channel, wang2020channel, elbir2020deep, liu2020deepdenoising}, we model the channel estimation problem in IRS-MC systems as a denoising problem and propose a CNN-based deep residual network (CDRN) to implicitly learn the residual noise for recovering the channel coefficients from the noisy pilot-based observations.

\item Specifically, to exploit both the spatial features of the noisy channel matrices and the additive nature of the noise simultaneously, a CNN denoising block equipped with an element-wise subtraction structure is designed for CDRN.
    Inheriting from the superiorities of CNN and DRN in feature extraction \cite{liu2019deep, liu2020deep} and denoising \cite{lxm2020deepresidual}, the proposed CDRN could further improve the estimation accuracy.

\item Finally, according to the MMSE criterion, a CDRN-based MMSE (CDRN-MMSE) estimator is derived in terms of Bayesian philosophy and the simulation results show that the performance of the proposed method approaches that of the optimal MMSE estimator requiring the computation of a prior probability density function (PDF) of the cascaded channel.
\end{itemize}

\emph{Organization}: Section \Rmnum{2} introduces the system model of the IRS-MC system.
In Section \Rmnum{3}, we model the channel estimation as a denoising problem, develop a CDRN architecture for channel estimation, and design the CDRN-based estimation algorithm.
Section \Rmnum{4} presents the numerical results to evaluate the estimation performance proposed method. Finally, we conclude the work of this paper in Section \Rmnum{5}.

\emph{Notations}:
Superscripts $H$ and $T$ denote the conjugate transpose and the transpose, respectively. Terms $\mathbb{C}$ and $\mathbb{R}$ represent the sets of complex numbers and real numbers, respectively.
${\mathcal{CN}}( \bm{\mu},\mathbf{\Sigma} )$ denotes the circularly symmetric complex Gaussian (CSCG) distribution where $\bm{\mu}$ and $\mathbf{\Sigma}$ are the mean vector and the covariance matrix, respectively. ${\mathbf{I}}_p$ is the $p$-by-$p$ identity matrix and ${\mathbf{0}}$ is a zero vector.
$(\cdot)^{-1}$ represents the matrix inverse.
$\mathrm{Re}\{\cdot\}$ and $\mathrm{Im}\{\cdot\}$ denote the extractions of the real part and the imaginary part of a complex-valued matrix, respectively.
$\mathrm{diag}(\cdot)$ denotes the construction of a diagonal matrix based on an input vector and $\mathrm{tr}[\cdot]$ represents the trace of a matrix. $\|\cdot\|_F$ is the Frobenius norm of a matrix. $E(\cdot)$ indicates the statistical expectation operation.


\begin{figure}[t]
  \centering
  \includegraphics[width=0.9\linewidth]{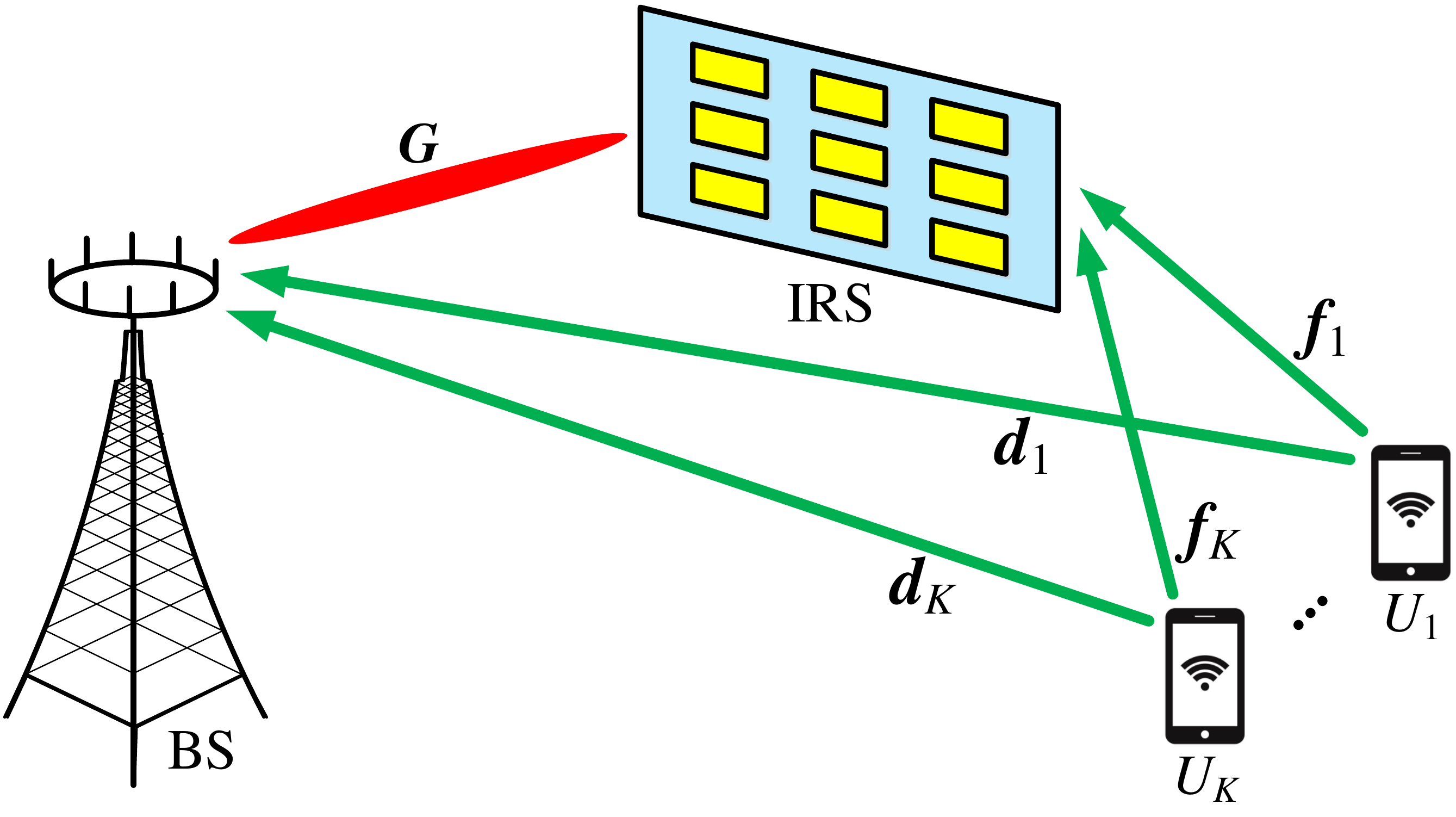}
  \caption{ The uplink of the considered IRS-MC system. }\label{Fig:uplink scenario}
\end{figure}

\section{System Model}
In this paper, an IRS-MC system adopting a time division duplex (TDD) protocol is considered, where an IRS equipped with $N$ passive reflecting elements is deployed to assist the communication between one base station (BS) and $K$ users, as shown in Fig. \ref{Fig:uplink scenario}. The BS and each user are equipped with an $M$-element antenna array and a single-antenna, respectively.
Due to the channel reciprocity property in TDD systems, the downlink CSI could be acquired from the uplink channel estimation in the IRS-MC system.
We focus on the uplink communication, where $U_k$, $k \in \{1,2,\cdots,K\}$, denotes the $k$-th user. The channels of the $U_k$-IRS link, the IRS-BS link, and the $U_k$-BS link, are represented by $\mathbf{f}_k \in \mathbb{C}^{N \times 1}$, $\mathbf{G} \in \mathbb{C}^{M \times N}$, and $\mathbf{d}_k \in \mathbb{C}^{M \times 1}$, respectively.
Note that each element at IRS combines all the arriving signals and reflects them to the BS behaving as a single point source, thus the reflecting link $U_k$-IRS-BS can be regarded as a dyadic backscatter channel \cite{gong2020towards}.
Let us denote the phase-shift matrix by $\mathbf{R}=\mathrm{diag}(\mathbf{r}) \in \mathbb{C}^{N \times N}$ with $\mathbf{r}=[\beta e^{j\varphi_1},\beta e^{j\varphi_2},\cdots,\beta e^{j\varphi_N}]^T$, where $0 \leq \beta \leq 1$ and $0 \leq \varphi_n \leq 2\pi$ are the amplitude and the phase shift at the $n$-th, $n \in \{1,2,\cdots,N\}$, element of the IRS, respectively.
In this case, the channel response of $U_k$-IRS-BS link can be expressed as $\mathbf{G}\mathbf{R}\mathbf{f}_k \in \mathbb{C}^{M \times 1}$. Considering that $\mathrm{diag}(\mathbf{r})\mathbf{f}_k=\mathrm{diag}(\mathbf{f}_k)\mathbf{r}$, the channel of $U_k$-IRS-BS link can be expressed as $\mathbf{G}\mathrm{diag}(\mathbf{r})\mathbf{f}_k = \mathbf{G}\mathrm{diag}(\mathbf{f}_k)\mathbf{r}$.
Therefore, the channel estimation objective is to estimate \vspace{-0.1cm}
\begin{equation}\label{H_k}
  \mathbf{H}_k = [\mathbf{d}_k,\mathbf{B}_k] \in \mathbb{C}^{M \times (N + 1)}, \forall k, \vspace{-0.1cm}
\end{equation}
where \vspace{-0.1cm}
\begin{equation}\label{B_k}
  \mathbf{B}_k = \mathbf{G}\mathrm{diag}(\mathbf{f}_k) \in \mathbb{C}^{M \times N}, \forall k, \vspace{-0.1cm}
\end{equation}
is a cascaded channel.


\begin{figure}[t]
  \centering
  \includegraphics[width=\linewidth]{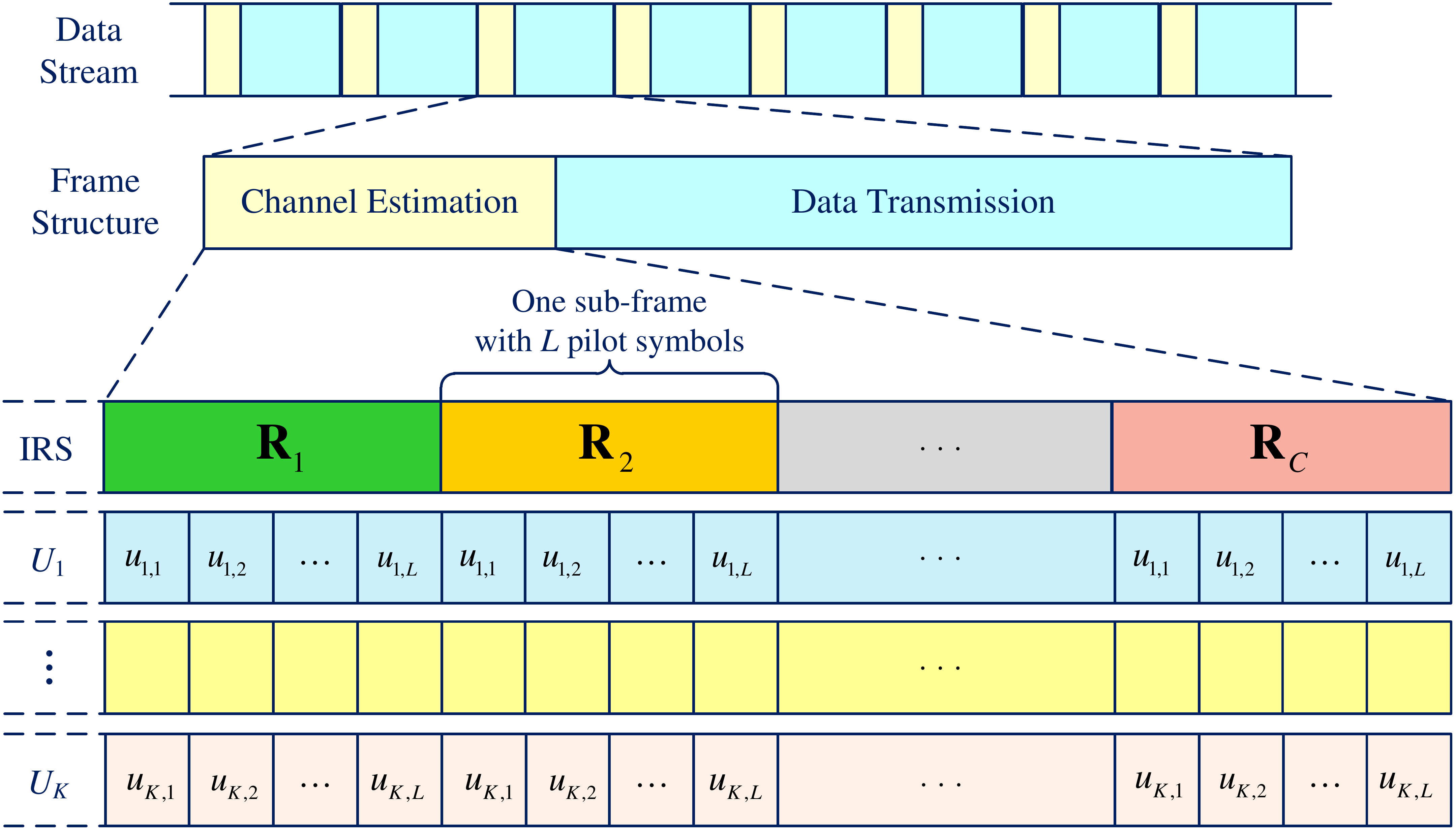} \vspace{-0.2cm}
  \caption{ Channel estimation protocol for the considered IRS-MC system. }\label{Fig:channel estimation protocol}\vspace{-0.6cm}
\end{figure}

In the following, we design a communication protocol for channel estimation in the considered IRS-MC system, as shown in Fig. \ref{Fig:channel estimation protocol}. The data stream consists of multiple identical frames and each frame consists of a channel estimation phase and a data transmission phase. The work of this paper mainly focuses on the channel estimation phase to estimate $\mathbf{H}_k$ as defined in (\ref{H_k}).
During the channel estimation phase, the IRS needs to generate $C$, $C \geq N + 1$, different reflection patterns via $C$ different phase-shift matrices, denoted by \vspace{-0.1cm}
\begin{equation}\label{Lambda}
  \mathbf{\Lambda} = [\mathbf{R}_1,\mathbf{R}_2,\cdots,\mathbf{R}_C]^{T}. \vspace{-0.1cm}
\end{equation}
Here, $\mathbf{R}_c = \mathrm{diag}(\mathbf{r}_c)$ represents the $c$-th, $c \in \{1,2,\cdots,C\}$, phase-shift matrix with $\mathbf{r}_c = [\beta_c e^{j\varphi_{c,1}},\beta_c e^{j\varphi_{c,2}},$ $\cdots,\beta_c e^{j\varphi_{c,N}}]$, where $\beta_c \in [0,1]$ and $\varphi_{c,n} \in [0,2\pi]$ are the amplitude and the phase shift of the $n$-th element of the IRS under the $c$-th phase-shift matrix, respectively.
For each $U_k$, $\forall k$, a pilot sequence $\mathbf{u}_k = [u_{k,1},u_{k,2},\cdots,u_{k,L}]^{T}$  with a length of $L$, $L \geq K$, is adopted for each IRS phase-shift matrix, where $u_{k,l}$ denotes the $l$-th, $l \in \{1,2,\cdots,L\}$, pilot symbol for $U_k$.
To distinguish different users, each two pilot sequences should be orthogonal, i.e., $\mathbf{u}_a^{H}\mathbf{u}_b = 0$ and $\mathbf{u}_k^{H}\mathbf{u}_k = \mathcal{P}L$, where $\mathcal{P}$ is the power of each pilot symbol, $a,b \in \{1,2,\cdots,K\}$, and $a \neq b$.
As shown in Fig. \ref{Fig:channel estimation protocol}, there are $C$ sub-frames in the channel estimation phase and each sub-frame consists of $L$ pilot symbols.
Specifically, the phase-shift matrix in IRS keeps unchanged within one sub-frame and it switches to different phase-shift matrices for different sub-frames. In this case, $U_k$ can send its identity pilot sequence $\mathbf{u}_k$ for each sub-frame.
Therefore, the received pilot signal vectors during the $c$-th sub-frame at the BS can be expressed as \vspace{-0.2cm}
\begin{align}
  \mathbf{S}_c = \sum\limits_{k = 1}^K \mathbf{H}_k\mathbf{p}_c\mathbf{u}_k^{H} + \mathbf{V}_c, \vspace{-0.2cm} \label{S_c}
\end{align}
Here, $\mathbf{S}_c = [\mathbf{s}_{c,1},\mathbf{s}_{c,2},\cdots,\mathbf{s}_{c,L}]\in \mathbb{C}^{M\times L}$ where $s_{c,l}\in \mathbb{C}^{M\times 1}$ denotes the $l$-th received pilot signal in the $c$-th sub-frame. $\mathbf{p}_c = [1,\mathbf{r}_c]^T \in \mathbb{C}^{(N + 1) \times 1}$. In addition, $\mathbf{V}_c = [\mathbf{v}_{c,1},\mathbf{v}_{c,2},\cdots,\mathbf{v}_{c,L}]$ where $\mathbf{v}_{c,l} \in \mathbb{C}^{M \times 1}$ is the $l$-th noise vector at the BS in the $c$-th sub-frame. Generally, we assume $\mathbf{v}_{c,l}$ to be a CSCG random vector with $\mathbf{v}_{c,l} \sim \mathcal{CN}(\mathbf{0},\sigma_v^2\mathbf{I}_M)$, where $\sigma_v^2$ is the noise variance of each antenna at the BS.
Taking advantage of the orthogonal property of the pilot sequences, we have \vspace{-0.1cm}
\begin{equation}\label{x_ck}
   {\mathbf{x}}_{c,k} = \mathbf{H}_k\mathbf{p}_c + {\mathbf{z}}_{c,k}. \vspace{-0.1cm}
\end{equation}
Here, ${\mathbf{x}}_{c,k} = \frac{1}{\mathcal{P}L}\mathbf{S}_c\mathbf{u}_k \in \mathbb{C}^{M \times 1}$ denotes the received signal vector at the BS from $U_k$ in the $c$-th sub-frame. Correspondingly, ${\mathbf{z}}_{c,k} = \frac{1}{\mathcal{P}L}\mathbf{V}_{c}\mathbf{u}_k \in \mathbb{C}^{M \times 1}$ is the noise component with ${\mathbf{z}}_{c,k} \sim \mathcal{CN}(\mathbf{0},\sigma_z^2\mathbf{I}_M)$, where $\sigma_z^2=\frac{1}{\mathcal{P}L}\sigma_v^2$.
After receiving $C$ subframes at the BS, we have \vspace{-0.1cm}
\begin{equation}\label{X_k}
  \mathbf{X}_k = \mathbf{H}_k\mathbf{P} + \mathbf{Z}_k, \vspace{-0.1cm}
\end{equation}
where $\mathbf{X}_k = [\mathbf{x}_{1,k},\mathbf{x}_{2,k},\cdots,\mathbf{x}_{C,k}]\in \mathbb{C}^{M \times C}$ is the result of stacking ${\mathbf{x}}_{c,k}$ from all the $C$ sub-frames,   $\mathbf{P} = [\mathbf{p}_{1},\mathbf{p}_{2},\cdots,\mathbf{p}_{C}] \in \mathbb{C}^{(N+1) \times C}$, and $\mathbf{Z}_k = [\mathbf{z}_{1,k},\mathbf{z}_{2,k},\cdots,\mathbf{z}_{C,k}] \in \mathbb{C}^{M \times C}$.
Generally, $\mathbf{P}$ can be designed as a discrete Fourier transform (DFT) form \cite{jensen2020optimal}, i.e., \vspace{-0.1cm}
\begin{equation}\label{P}
  \mathbf{P} = \left[
  \begin{matrix}
  1      & 1      & \cdots & 1      \\
 1      & W_C      & \cdots & W_C^{C-1}      \\
 \vdots & \vdots & \ddots & \vdots \\
 1      & W_C^N      & \cdots & W_C^{N(C - 1)}      \\
 \end{matrix}
 \right] \in \mathbb{C}^{(N+1) \times C}, \vspace{-0.1cm}
\end{equation}
where $W_C = e^{j2\pi/C}$ and $\mathbf{P}\mathbf{P}^H = C\mathbf{I}_{N + 1}$. Thus, the reflection pattern $\mathbf{R}_c = \mathrm{diag}(\mathbf{r}_c)$ in (\ref{Lambda}) can be determined based on (\ref{P}).

According to (\ref{X_k}) and (\ref{P}), the objective of channel estimation in IRS-MC is to recover $\mathbf{H}_k$, $\forall k$, by exploiting $\mathbf{X}_k$ and $\mathbf{P}$.
Next, we first introduce two commonly adopted methods as benchmarks, i.e., the LS method and the LMMSE method.

(a)~\emph{LS Channel Estimator}:

If no prior knowledge of channel is available, $\mathbf{H}_k$ is assumed to be an unknown but deterministic constant. In this case, the LS estimator \cite{biguesh2006training} is a practical method, i.e., \vspace{-0.1cm}
\begin{equation}\label{LSv1}
  \tilde{\mathbf{H}}_k^{\mathrm{LS}} = \mathbf{X}_k\mathbf{P}^\dag, \vspace{-0.1cm}
\end{equation}
where $\tilde{\mathbf{H}}_k^{\mathrm{LS}}$ denotes the estimated $\mathbf{H}_k$ adopted the LS estimator and $\mathbf{P}^\dag = \mathbf{P}^H(\mathbf{P}\mathbf{P}^H)^{-1}$ is the pseudoinverse of $\mathbf{P}$.

(b)~\emph{LMMSE Channel Estimator}:

If the statistical channel correlation matrix $\mathbf{R}_{\mathbf{H}_k} = E(\mathbf{H}_k^H\mathbf{H}_k)$ is available, $\mathbf{H}_k$ can be assumed to be a random variable \cite{biguesh2006training, liu2019maximum}.
In this case, the LMMSE estimator is developed to explore the statistical features of channel to further improve the estimation accuracy and its expression is given by \cite{biguesh2006training}
\begin{equation}\label{LMMSE}
  \tilde{\mathbf{H}}_k^{\mathrm{LMMSE}} = \mathbf{X}_k(\mathbf{P}^H\mathbf{R}_{\mathbf{H}_k}\mathbf{P} + M\sigma_z^2\mathbf{I}_C)^{-1}\mathbf{P}^{H}\mathbf{R}_{\mathbf{H}_k},
\end{equation}
where $\tilde{\mathbf{H}}_k^{\mathrm{LMMSE}}$ is the estimated $\mathbf{H}_k$ by LMMSE estimator.


\begin{figure*}[t]
  \centering
  \includegraphics[width=\linewidth]{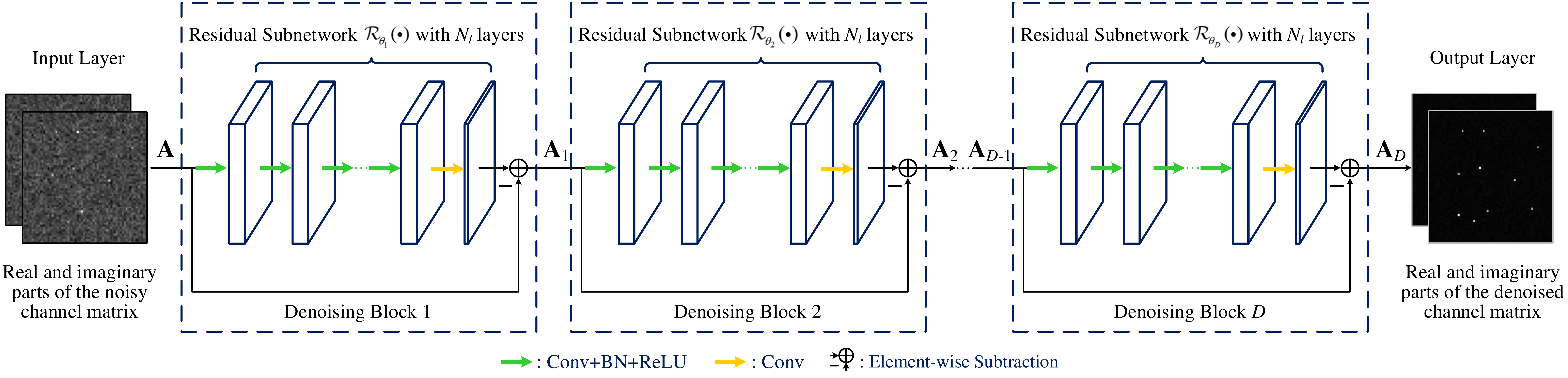}
  \caption{ The network architecture of the proposed CDRN for channel estimation in the IRS-MC system. }\label{Fig:CDRN_architecture}
\end{figure*}

Note that if the channel follows the Rayleigh fading model and the statistical channel correlation matrix is available, the LMMSE estimator is equivalent to the optimal MMSE estimator \cite{kay1993fundamentals}, i.e., the LMMSE estimator achieves the optimal performance.
However, for IRS-MC systems, $\mathbf{H}_k$ contains a cascaded channel, $\mathbf{B}_k$ in (\ref{B_k}), which generally does not follow the Rayleigh fading model.
In this case, the LMMSE estimator behaves differently compared with the optimal MMSE estimator as the former can not achieve the optimal estimation performance. Meanwhile, implementing the optimal MMSE estimator is intractable in practice due to its intensive computational cost and the unknown end-to-end channel distribution.

To further improve the channel estimation performance, in the next section, we will retain the MSE criterion and adopt a data driven approach to design a DL-based estimation method.

\section{CDRN for Channel Estimation}
Note that the additive noise in the system model hinders the accurate recovery of the channel coefficients.
In this section, the channel estimation is modeled as a denoising problem and we develop a CDRN to implicitly learn the residual noise from the noisy observations for recovering the channel coefficients.
Taking advantage of the powerful capabilities of CNN and DRN in feature extraction and denoising, the proposed CDRN could further improve the estimation accuracy.
In the following, we will introduce the derived denoising model, the developed CDRN architecture, and the designed CDRN-based channel estimation algorithm, respectively.

\subsection{Denoising Model for Channel Estimation}
Note that since the LS estimator requires no prior knowledge of data, it has been widely adopted in practice for its convenience of implementation.
According to \cite{biguesh2006training}, the mean square error of the LS estimation is \vspace{-0.3cm}
\begin{equation}\label{error_LS}
\varepsilon_{\mathrm{LS}} = E(\|\mathbf{H}_k - \tilde{\mathbf{H}}_k^{\mathrm{LS}}\|_F^2) = \frac{M\sigma_z^2}{(N + 1)C}. \vspace{-0.3cm}
\end{equation}
It is obvious that $\varepsilon_{\mathrm{LS}}$ is a function of the noise power.
Therefore, we can adopt the estimated channel value by exploiting the LS estimator as the coarse estimated value and thus the channel estimation for IRS-MC systems can be modeled as a denoising problem: recovering $\mathbf{H}_k$ from a noisy observation
\begin{equation}\label{LSv2}
  \tilde{\mathbf{X}}_k = {\mathbf{H}}_k + \tilde{\mathbf{Z}}_k,
\end{equation}
where $\tilde{\mathbf{X}}_k = \mathbf{X}_k\mathbf{P}^\dag \in \mathbb{C}^{M \times (N + 1)}$ denotes the noisy observation obtained by the LS estimator and $\tilde{\mathbf{Z}}_k = \mathbf{Z}_k\mathbf{P}^\dag \in \mathbb{C}^{M \times (N + 1)}$ is the noise component. Note that $\mathbf{H}_k$ contains a cascaded channel $\mathbf{B}_k$ whose prior PDF is generally not a Gaussian PDF, thus, the data model in (\ref{LSv2}) is not a Bayesian general linear model \cite{kay1993fundamentals}.
In this case, it is intractable to derive the closed-form of the optimal MMSE estimator. Therefore, we will adopt a data-driven approach to develop a CDRN for channel estimation in IRS-MC systems in the following section.

\subsection{CDRN Architecture}
The CDRN architecture consists of one input layer, $D$ denoising blocks, and one output layer, as shown in Fig. \ref{Fig:CDRN_architecture}.
Specifically, $D$ denoising blocks are successively cascaded to gradually enhance the denoising performance and all denoising blocks have an identical structure.
The hyperparameters of CDRN are summarized in Table I and we will introduce each layer of CDRN in the following.

\emph{a)~Input Layer}:
Note that since the noisy channel matrix is a complex-valued matrix, we adopt two neural network channels to process the real part and imaginary part of the noisy channel matrix, respectively. In this case, denote by $\mathbf{A} \in \mathbb{R}^{M \times (N+1) \times 2}$ the input of CDRN, we have
\begin{equation}\label{I_input}
  \mathbf{A} = \mathcal{F}([ \mathrm{Re}\{\tilde{\mathbf{X}}_k\}, \mathrm{Im}\{\tilde{\mathbf{X}}_k\} ]),
\end{equation}
where $\mathcal{F}(\cdot)\!\!:\!\!\mathbb{R}^{M \times (2N+2)}\mapsto\mathbb{R}^{M \times (N+1) \times 2}$ denotes the mapping function and $\tilde{\mathbf{X}}_k$ is the noisy observation based on the LS estimator, as defined in (\ref{LSv2}).

\emph{b)~Denoising Blocks}:
As shown in Fig. \ref{Fig:CDRN_architecture}, there are $D$ identical denoising blocks in CDRN, which are adopted to gradually enhance the denoising performance. Each denoising block consists of a $N_l$-layer residual subnetwork and an element-wise subtraction. The detailed hyperparameters are shown in Table I.
For the first $N_l-1$ layers of the residual subnetwork, we adopt the ``Conv+BN+ReLU'' operations denoted by the green arrows for each layer.
The ``Conv+BN+ReLU'' operation refers to a successively cascaded combination of convolution (Conv)-to-batch normalization (BN)-to-rectified linear unit (ReLU).
Specifically, the Conv and the ReLU are adopted jointly to explore the spatial features of channel matrices and the BN is added between them to improve the network stability and the network training speed.
For the last layer of the residual subnetwork, we adopt a Conv operation denoted by the yellow arrow to combine the extracted features to construct the residual noise matrix for the subsequent element-wise subtraction.
Finally, an element-wise subtraction is adopted to exploit the additive nature of the noise for denoising the noisy channel matrix.

\begin{table}[t]
\normalsize
\caption{Hyperparameters of the proposed CDRN}
\vspace{-0.3cm}
\centering
\small
\renewcommand{\arraystretch}{1.2}
\begin{tabular}{c c c}
  \hline
   \multicolumn{3}{l}{\textbf{Input}: Noisy channel matrix ($M \times (N+1) \times 2$)} \\
  \hline
   \multicolumn{3}{l}{\textbf{Denoising Block}: (There are $D$ identical denoising blocks)}  \\
  \hspace{0.6cm} \textbf{Layers} & \textbf{Operations} & \textbf{Filter Size}   \\
  \hspace{0.6cm} 1 & Conv + BN + ReLU  & $ 64 \times ( 3 \times 3 \times 2 ) $   \\
  \hspace{0.6cm} $ 2 \sim N_l-1 $ & Conv + BN + ReLU &  $ 64 \times ( 3 \times 3 \times 64 ) $   \\
  \hspace{0.6cm} $N_l$ & Conv & $ 2 \times ( 3 \times 3 \times 64 ) $  \\
  \hline
   \multicolumn{3}{l}{\textbf{Output}: Denoised channel matrix ($M \times (N+1) \times 2$)} \\
  \hline
\end{tabular}
\end{table}

\begin{figure*}[t]
  \centering
  \includegraphics[width=0.9\linewidth]{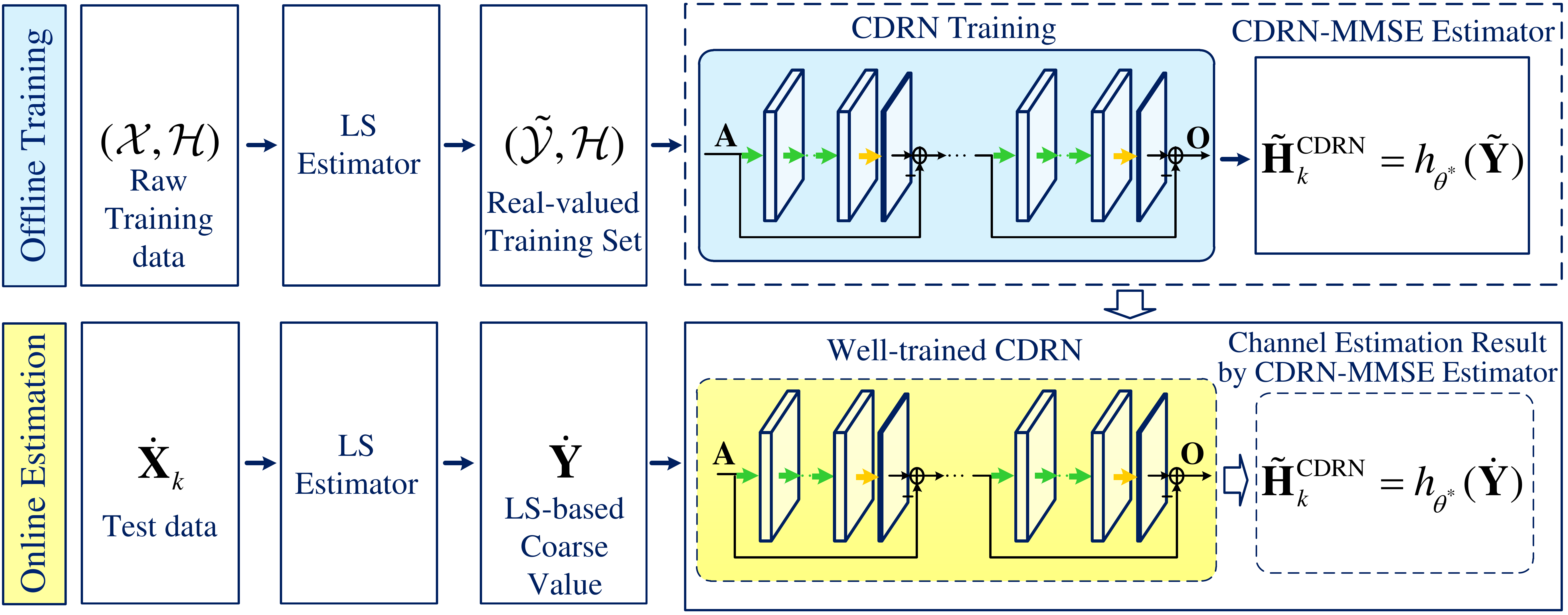}
  \caption{ The developed CDRN-based framework for channel estimation in the considered IRS-MC system. }\label{Fig:CDRN_framework}
\end{figure*}

Let $\mathcal{R}_{\theta_d}(\cdot)$ denote the function expression of the $d$-th, $d \in \{1,2,\cdots,D\}$, residual subnetwork, where $\theta_d$ is the network parameters.
In this case, we can formulate the $d$-th denoising block as
\begin{equation}\label{I_d}
  \mathbf{A}_d = \mathbf{A}_{d-1} - \mathcal{R}_{\theta_d}(\mathbf{A}_{d-1}), \forall d,
\end{equation}
where $\mathbf{A}_0 = \mathbf{A}$, $\mathbf{A}_{d-1}$ and $\mathbf{A}_d$ denote the input and the output of the $d$-th denoising block, respectively.

\emph{c)~Output Layer}:
The output layer is the output result of the $D$-th denoising block which is given by
\begin{equation}\label{I_D}
  \mathbf{A}_D = \mathbf{A}_{D-1} - \mathcal{R}_{\theta_D}(\mathbf{A}_{D-1}).
\end{equation}


Based on (\ref{I_d}) and (\ref{I_D}), the output of the proposed CDRN can be formulated as
\begin{equation}\label{h_theta}
  h_{\theta}(\mathbf{A}) = \mathbf{A} - \sum\limits_{d = 1}^{D}\mathcal{R}_{\theta_d}\left(\mathbf{A}_{d-1}\right),
\end{equation}
where $h_{\theta}(\cdot)$ represents the expression of CDRN with parameters $\theta = \{\theta_1,\theta_2,\cdots,\theta_D\}$ and $\sum_{d = 1}^{D}\mathcal{R}_{\theta_d}\left(\mathbf{A}_{d-1}\right)$ denotes the residual noise component.
Therefore, the obtained channel matrix is a denoising result, i.e., the element-wise subtraction between the observed noisy channel matrix $\mathbf{A}$ and the residual noise component $\sum_{d = 1}^{D}\mathcal{R}_{\theta_d}\left(\mathbf{A}_{d-1}\right)$.

\subsection{CDRN-based Channel Estimation Algorithm}
Based on the designed CDRN, we propose a CDRN-based channel estimation algorithm, as shown in Fig. \ref{Fig:CDRN_framework}, which consists of offline training and online estimation phases.

(a)~\emph{Offline Training Phase}:
According to the denoising model in (\ref{LSv2}), the training set can be formulated as
\begin{equation}\label{LS_trainingset_CDRN}
  (\tilde{\mathcal{Y}},\mathcal{H}) = \{ ( \tilde{\mathbf{Y}}_k^{(1)},\mathbf{H}_k^{(1)} ), ( \tilde{\mathbf{Y}}_k^{(2)},\mathbf{H}_k^{(2)} ), \cdots, ( \tilde{\mathbf{Y}}_k^{(N_t)},\mathbf{H}_k^{(N_t)} )  \},
\end{equation}
where $\tilde{\mathbf{Y}}_k^{(i)} = \mathcal{F}([\mathrm{Re}\{\tilde{\mathbf{X}}_k^{(i)}\},\mathrm{Im}\{\tilde{\mathbf{X}}_k^{(i)}\}]) \in \mathbb{R}^{M \times (N+1) \times 2}$ and $\mathbf{H}_k^{(i)}$ denote the input and the label of the $i$-th, $i \in \{1,2,\cdots, N_t\}$, training example of $(\tilde{\mathcal{Y}},\mathcal{H})$, respectively.

\begin{table}[t]
\small
\centering
\begin{tabular}{l}
\toprule[1.8pt] \vspace{-0.3cm}\\
\hspace{-0.1cm} \textbf{Algorithm 1} {CDRN-based Channel Estimation Algorithm} \\
\toprule[1.8pt] \vspace{-0.3cm}\\
\textbf{Initialization:} $i_t = 0$, real-valued training set $(\tilde{\mathcal{Y}},\mathcal{H})$ \\
\textbf{Offline Training Phase:} \\
1:\hspace{0.75cm}\textbf{Input:} Training set $(\tilde{\mathcal{Y}},\mathcal{H})$\\
2:\hspace{1.1cm}\textbf{while} $i_t \leq I_t $ \textbf{do} \\
3:\hspace{1.6cm}Update $\theta$ by BP algorithm to minimize $J_{\mathrm{CDRN}}(\theta)$ \\
\hspace{1.8cm} $i_t = i_t + 1$  \\
4:\hspace{1.1cm}\textbf{end while} \\
5:\hspace{0.75cm}\textbf{Output}:  Well-trained CDRN ${h}_{\theta^*}( \cdot ) $\\
\textbf{Online Estimation Phase:} \\
6:\hspace{0.75cm}\textbf{Input:} Test data $\dot{\mathbf{Y}} = \mathcal{F}([ \mathrm{Re}\{\dot{\mathbf{X}}_k\mathbf{P}^\dag\}, \mathrm{Im}\{\dot{\mathbf{X}}_k\mathbf{P}^\dag\} ])$ \\
7:\hspace{1.1cm}\textbf{do} Channel Estimation using (\ref{CDRN_estimator}) \\
8:\hspace{0.75cm}\textbf{Output:} $\tilde{\mathbf{H}}_k^{\mathrm{CDRN}} = h_{\theta^*}(\dot{\mathbf{Y}})$. \vspace{0.2cm}\\
\bottomrule[1.8pt]
\end{tabular}
\end{table}

According to the MMSE criterion, the optimal MMSE estimator is obtained through minimizing the statistical Bayesian MSE. However, due to the limitation of the finite number of training examples, the statistical Bayesian MSE is not available and we can only adopt the empirical MSE as the cost function which is defined by
\begin{equation}\label{Jmse_CDRN}
  J_{\mathrm{CDRN}}(\theta) = \frac{1}{2N_t}\sum_{q=1}^{N_t} \left\|h_{\theta}(\tilde{\mathbf{Y}}_k^{(i)}) - \mathbf{H}_k^{(i)}\right\|_F^2.
\end{equation}
By minimizing (\ref{Jmse_CDRN}) through the backpropagation (BP) algorithm, we can finally obtain a well-trained CDRN, i.e., the CDRN-based MMSE (CDRN-MMSE) estimator:
\begin{equation}\label{CDRN_estimator}
  \tilde{\mathbf{H}}_k^{\mathrm{CDRN}} = h_{\theta^*}(\tilde{\mathbf{Y}}) = \tilde{\mathbf{Y}} - \sum\limits_{d = 1}^{D}\mathcal{R}_{\theta^*_d}\left(\tilde{\mathbf{Y}}_{d-1}\right).
\end{equation}
Here, $\tilde{\mathbf{H}}_k^{\mathrm{CDRN}}$ denotes the estimated channel matrix by CDRN. $\tilde{\mathbf{Y}}$ represents an arbitrary input matrix and $h_{\theta^*}(\cdot)$ denotes the well-trained CDRN with the well-trained network parameters $\theta^* = \{\theta^*_1,\theta^*_2,\cdots,\theta^*_D\}$. In addition, $\tilde{\mathbf{Y}}_{0} = \tilde{\mathbf{Y}}$ while $\tilde{\mathbf{Y}}_{d-1}$ and $\tilde{\mathbf{Y}}_{d}$ denote the input and the output of the $d$-th denoising block of CDRN as defined in (\ref{I_d}).

(b)~\emph{Online Estimation Phase}:
Given a test data $\dot{\mathbf{X}}_k$, we first adopt the LS estimator to obtain a coarse estimation result $\dot{\mathbf{X}}_k\mathbf{P}^\dag$ and then send it to the well-trained CDRN estimator to obtain a refined channel estimation result $  \tilde{\mathbf{H}}_k^{\mathrm{CDRN}} = h_{\theta^*}(\dot{\mathbf{Y}})$,
where $\dot{\mathbf{Y}} = \mathcal{F}([ \mathrm{Re}\{\dot{\mathbf{X}}_k\mathbf{P}^\dag\}, \mathrm{Im}\{\dot{\mathbf{X}}_k\mathbf{P}^\dag\} ])$ as defined in (\ref{I_input}).

(c)~\emph{CDRN-based Channel Estimation Algorithm Steps}:
Based on both the offline and online training phases, the proposed CDRN-based channel estimation algorithm is summarized in Algorithm 1, where $i_t$ is the iteration index and $I_t$ is the maximum iteration number of offline training.

\begin{figure}[t]
  \centering
  \includegraphics[width=3.6in,height=2.6in]{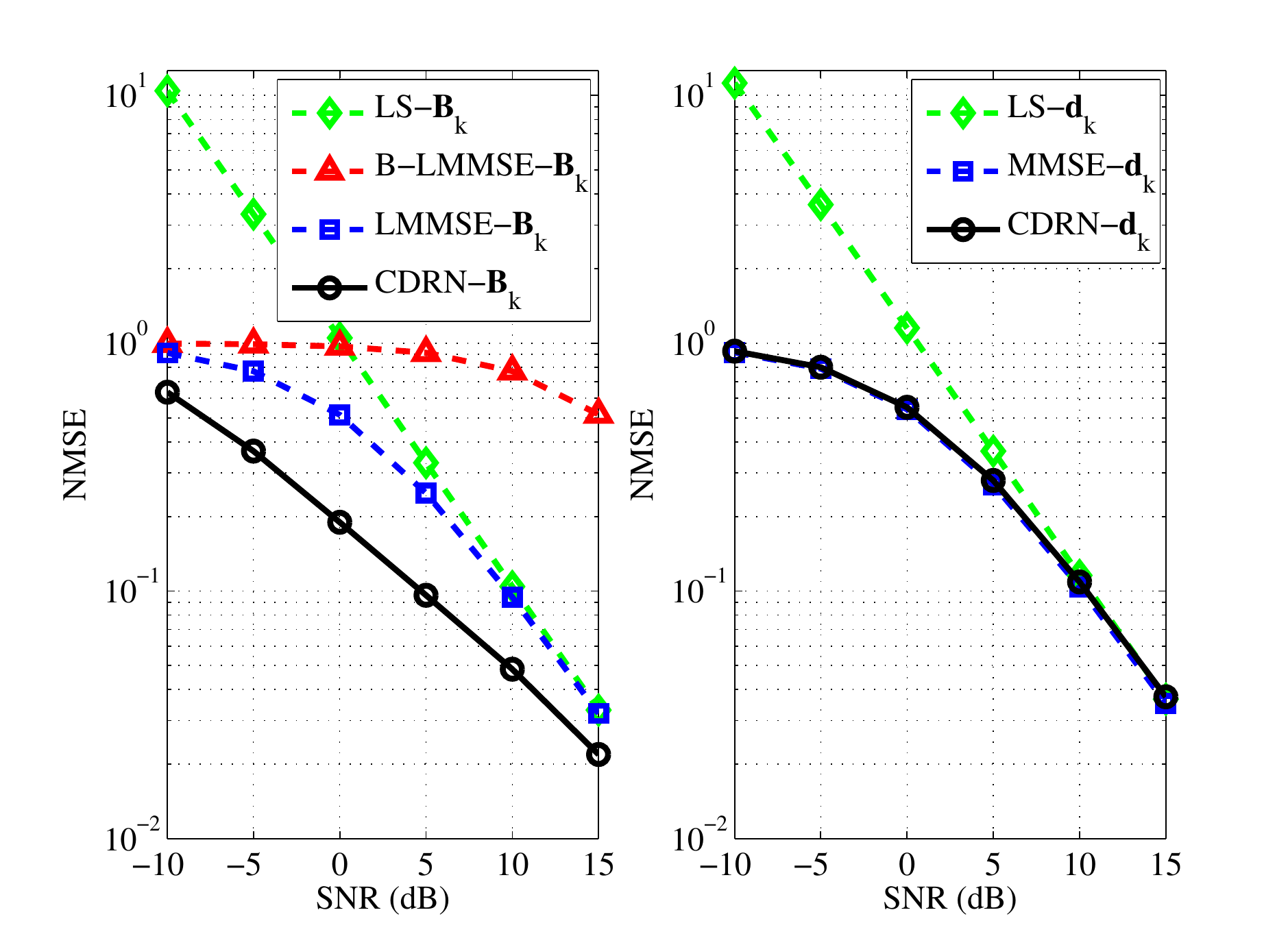} \\
  \hspace{0.4cm}{\scriptsize{(a) $\mathbf{B}_k$ for reflecting link. }} \hspace{1cm}{\scriptsize{(b) $\mathbf{d}_k$ for direct link. }}  \\
  \caption{NMSE performance under $M$ = 8, $N = 32$, and $C = 33$.}\label{Fig:NMSE_MC_SNR}
\end{figure}

\section{Numerical Results}
In this section, simulation results are presented to verify the efficiency of the proposed scheme.
In the simulation, an IRS-MC system is considered, as shown in Fig. \ref{Fig:uplink scenario}, where the BS is equipped with $M=8$ antennas, the IRS is with $N=32$ reflecting elements, and the number of single-antenna users is $K=6$.
Generally, a Rician channel model is assumed for all the links in the IRS-MC system.
In this case, the channel of IRS-BS link can be formulated as \cite{wu2019intelligent}
\begin{equation}\label{G_model}
  \mathbf{G} = \sqrt{\frac{\beta^{\mathrm{IB}}}{\beta^{\mathrm{IB}} + 1}}\bar{\mathbf{G}} + \sqrt{\frac{1}{\beta^{\mathrm{IB}} + 1}}\tilde{\mathbf{G}},
\end{equation}
where $\beta^{\mathrm{IB}}$ is the Rician factor of IRS-BS channel, and $\bar{\mathbf{G}}$ and $\tilde{\mathbf{G}}$ are the line-of-sight (LOS) and non-LOS (NLOS) components, respectively.
In addition, the path loss can be modeled as $\alpha^{\mathrm{IB}} = \alpha_0(\lambda^{\mathrm{IB}}/\lambda_0)^{-\gamma^{\mathrm{IB}}}$, where $\alpha_0 = -15~\mathrm{dB}$ is the path loss at the reference distance $\lambda_0 = 10~\mathrm{m}$ and $\gamma^{\mathrm{IB}}$ is the path loss exponent. Thus, the path loss factor can be taken into account by multiplying $\mathbf{G}$ in (\ref{G_model}) with $\sqrt{\alpha^{\mathrm{IB}}}$.
Similarly, the channels of other links can be generated via the same logic as defined in (\ref{G_model}) and the corresponding parameters are set as follows.
The distances of $U_k$-BS, IRS-BS, and $U_k$-IRS are $\lambda_k^{\mathrm{UB}} = 100$ m, $\lambda^{\mathrm{IB}} = 90$ m, and $\lambda^{\mathrm{UI}}_k = 16$ m.
The path loss exponents are $\gamma^{\mathrm{UB}}_k = 3.6$, $\gamma^{\mathrm{IB}} = 2.3$, and $\gamma^{\mathrm{UI}}_k = 2$. In addition, the Rician factors are set as $\beta^{\mathrm{UB}}_k = 0$, $\beta^{\mathrm{IB}} = 10$, and $\beta^{\mathrm{UI}}_k = 0$.
Specifically, a normalized MSE (NMSE) is selected as the performance metric of channel estimation \cite{liu2020deep}.
The SNR in the simulation results is set as the transmit SNR which is defined as $\mathrm{SNR} = \mathcal{P}/\sigma_z^2$.
For the proposed CDRN method, the number of training examples are set as $N_t=60,000$ and its hyperparameters used in the simulations are based on Table \Rmnum{1} with $D=3$.
In addition, each simulation point is the result of averaging over $100,000$ Monte Carlo realizations.

We first present the NMSE results of $\mathbf{B}_k$ and $\mathbf{d}_k$ in Fig. \ref{Fig:NMSE_MC_SNR}(a) and Fig. \ref{Fig:NMSE_MC_SNR}(b), respectively.
It is firstly noticed that the binary reflection controlled LMMSE (B-LMMSE) method \cite{mishra2019channel} performs worse than that of the DFT controlled methods (denoted by LS \cite{biguesh2006training}, LMMSE \cite{kay1993fundamentals}, and CDRN). This is because at each time slot, only one element is switched on for the B-LMMSE method, which leads to an insufficient received signal power for channel estimation.
In addition, it is shown that the proposed CDRN method outperforms both the LS method and the LMMSE method significantly, achieving the best performance among all the considered algorithms.
The reason is that the LMMSE estimator is a linear estimator. In contrast, the proposed method is a non-linear CDRN estimator, which can further improve the estimation performance by intelligently exploiting the non-linear spatial features of channels in a data driven approach.
On the other hand, in Fig. \ref{Fig:NMSE_MC_SNR}(b), the optimal MMSE method \cite{kay1993fundamentals} is equivalent to the LMMSE method since $\mathbf{d}_k$ is a Rayleigh fading channel.
It can be observed that the performance of the CDRN approaches that of the optimal MMSE method, which verifies the optimality of the proposed CDRN method.

\begin{figure}[t]
\centering
\subfigure[]{
\includegraphics[height=0.3\linewidth]{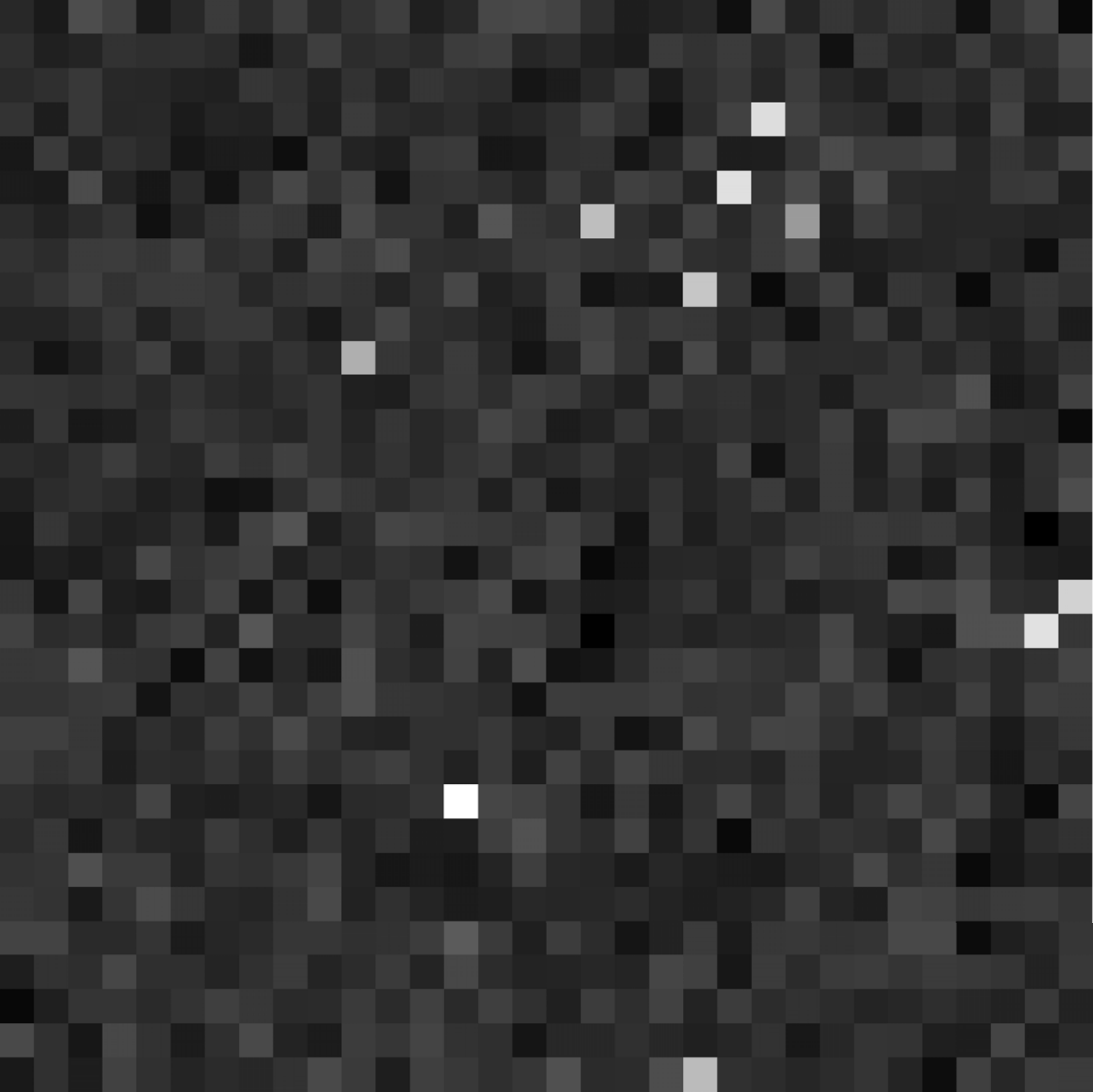}
\label{fig:Vis_Input}}
\,
\subfigure[]{
\includegraphics[height=0.3\linewidth]{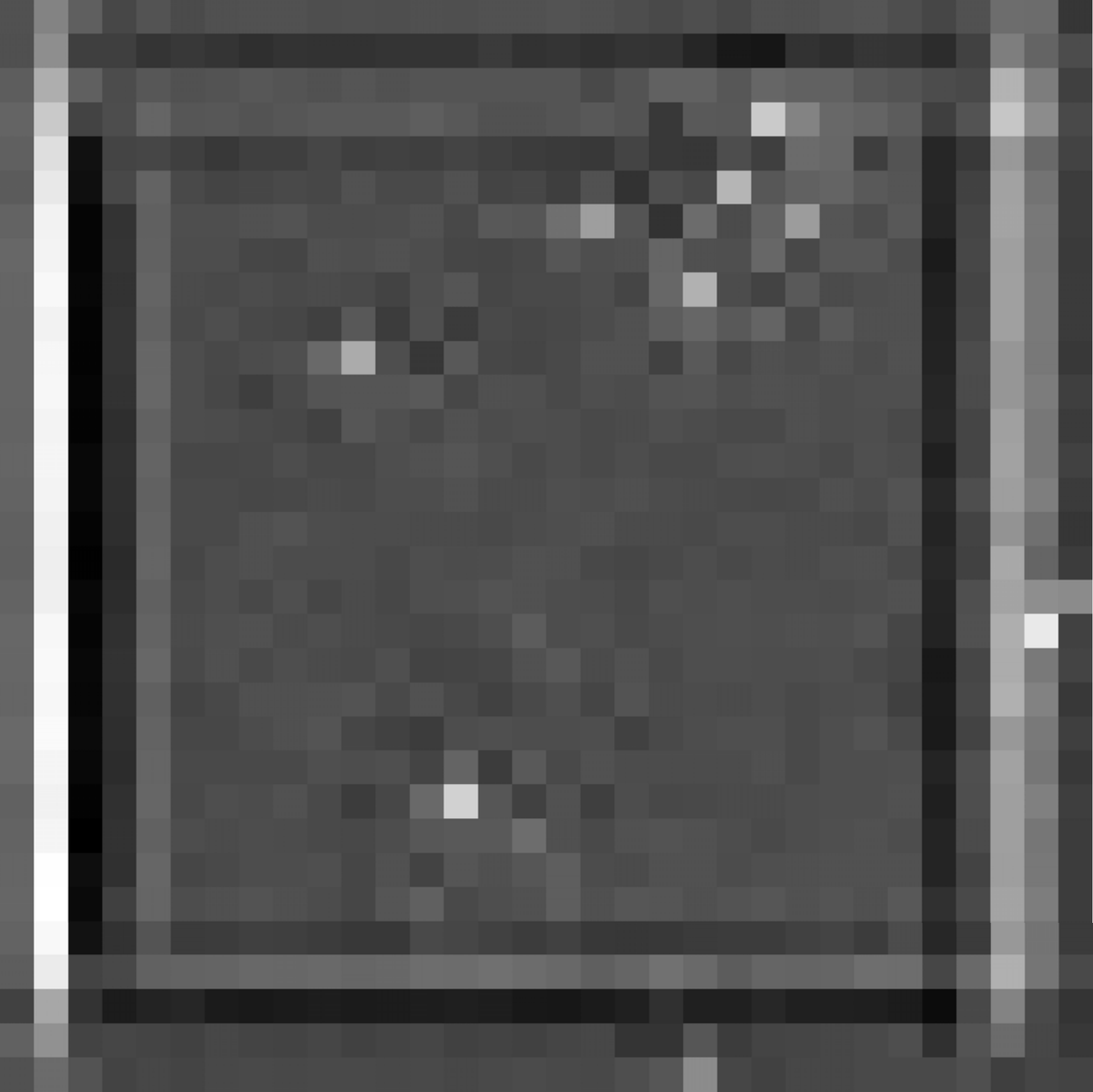}
\label{fig:Vis_B1}}
\,
\subfigure[]{
\includegraphics[height=0.3\linewidth]{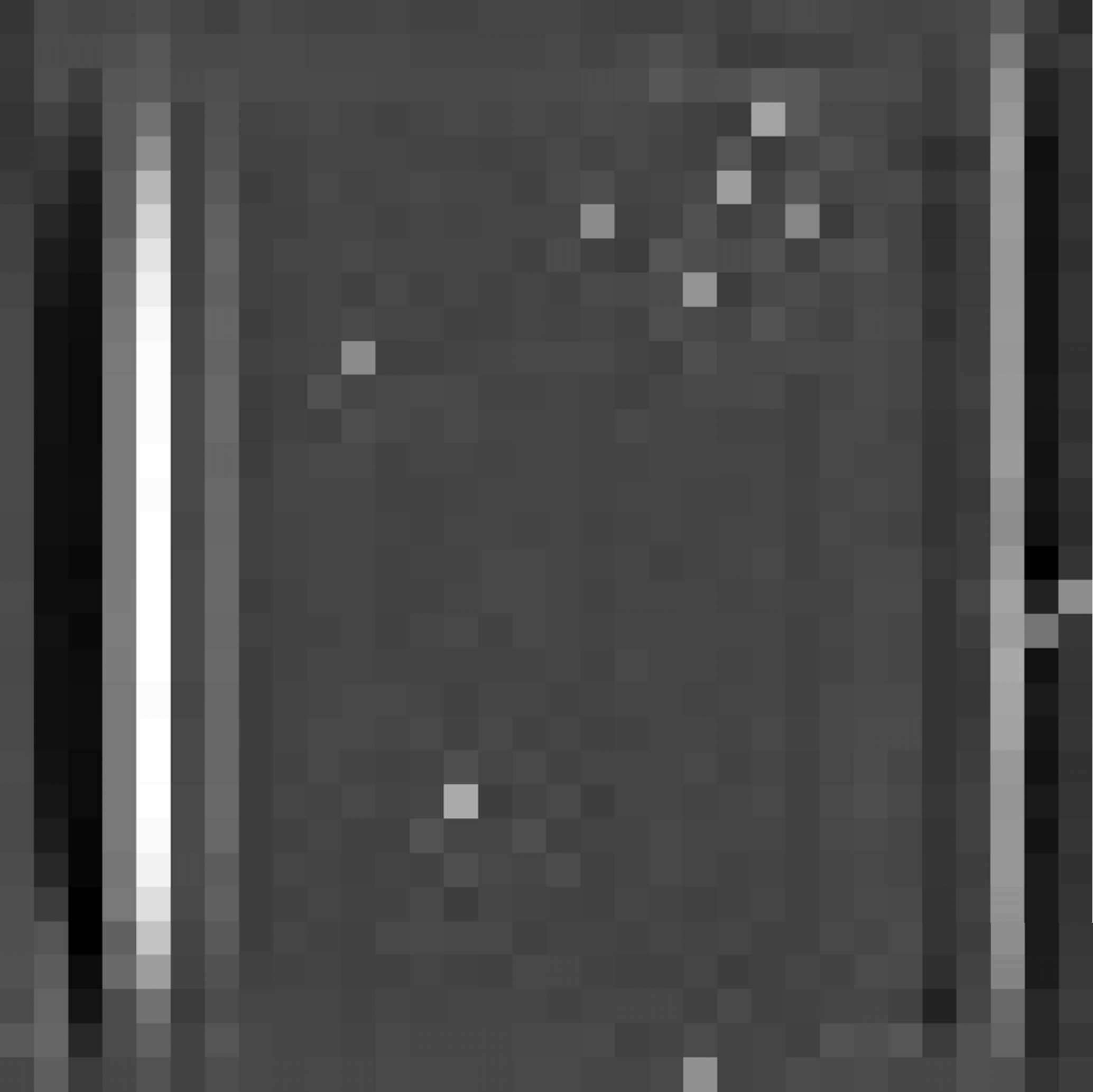}
\label{fig:Vis_B2}}
\,
\subfigure[]{
\includegraphics[height=0.3\linewidth]{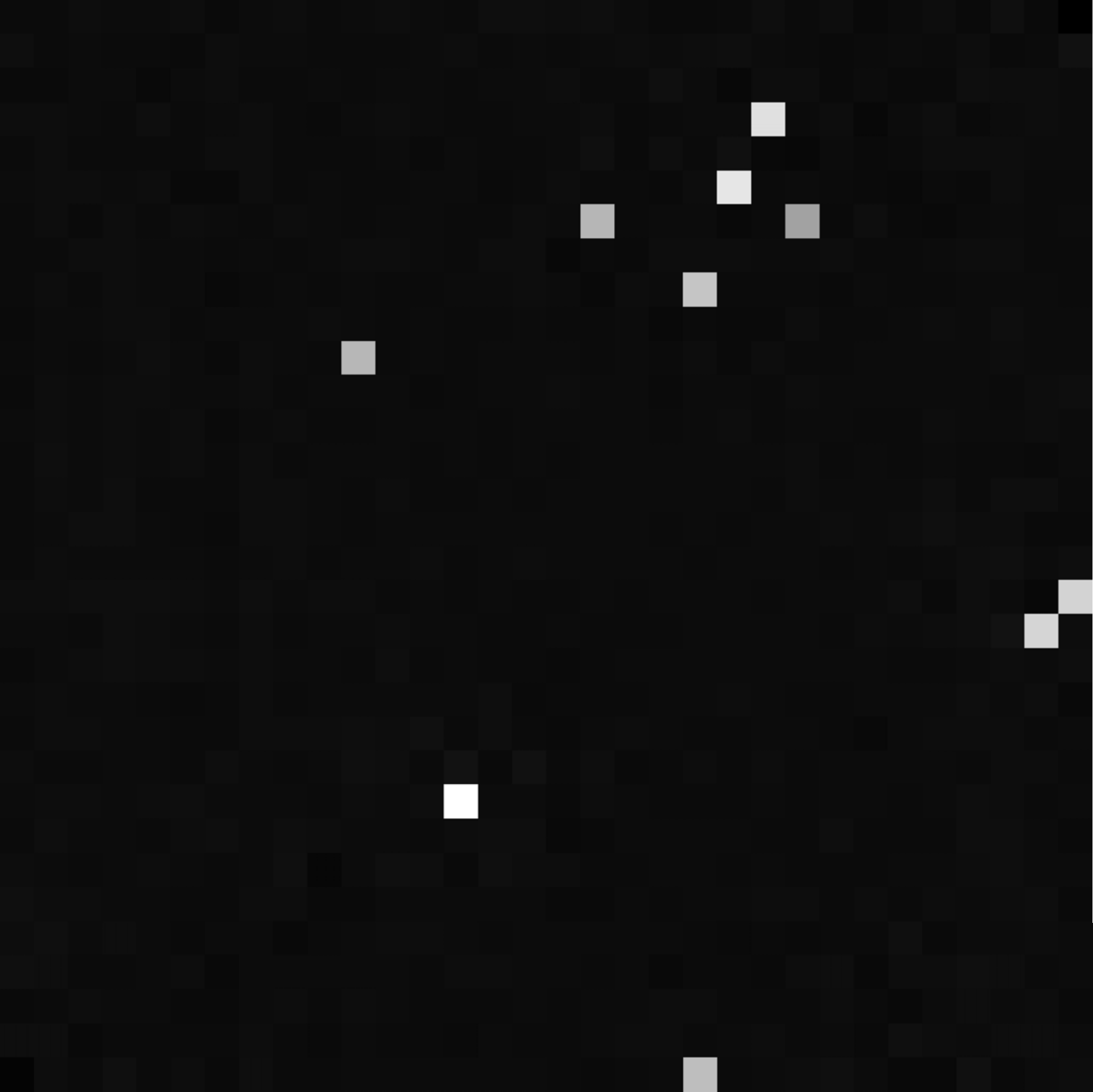}
\label{fig:Vis_B3}}
\caption{A visualization of denoising blocks of a well-trained CDRN with $D=3$ under SNR = 16 dB. (a) $\mathbf{A}$; (b) $\mathbf{A}_1$; (c) $\mathbf{A}_2$; (d) $\mathbf{A}_3$.}\label{Fig:VIS}
\end{figure}

To further understand the CDRN method, the visualizations of the outputs of each denoising block of a well-trained CDRN is presented in Fig. \ref{Fig:VIS}, where $(a),(b),(c)$, and $(d)$ are respectively the visualizations of the input of CDRN ($\mathbf{A}$), the output of the first, second, and last denoising blocks ($\mathbf{A}_1$, $\mathbf{A}_2$, and $\mathbf{A}_3$), as defined in Fig. \ref{Fig:CDRN_architecture}.
In Fig. \ref{fig:Vis_Input}, the ten distinguishable bright pixels are the noisy observations including interest channel coefficients while the remaining pixels are the noise samples.
Note that these cluttered noise pixels indicate a large noise variance.
Subsequently, it can be observed from Fig. \ref{fig:Vis_B1} and Fig. \ref{fig:Vis_B2} that the noise pixels of $\mathbf{A}_1$ and $\mathbf{A}_2$ become tidy progressively, i.e., the noise variance becomes small after the first and the second denoising blocks.
Finally, $\mathbf{A}_2$ is sent to the last denoising block and the visualization of the CDRN output is presented in Fig. \ref{fig:Vis_B3}.
We can find that the noise pixels have already been eliminated efficiently and thus a denoised channel matrix is finally obtained.
Therefore, Fig. \ref{Fig:VIS} indicates that the proposed CDRN method can achieve a satisfactory performance through a stage-by-stage denoising mechanism.

\section{Conclusion}
This paper developed a data-driven approach to address the channel estimation problem in IRS-MC systems.
In contrast to the existing channel estimation schemes, we modeled the channel estimation as a denoising problem and designed a CDRN to implicitly learn the residual noise for recovering the channel coefficients from the noisy observations.
Specifically, exploiting the powerful capabilities of CNN and DRN in feature extraction and denoising, a CNN denoising block with an element-wise subtraction structure is designed for CDRN, which further improve the estimation accuracy.
Finally, a CDRN-MMSE estimator is derived in terms of the MMSE criterion.
Simulation results showed that the performance of the proposed method approaches that of the optimal MMSE estimator requiring the PDF of the communication channel.

\bibliographystyle{ieeetr}

\setlength{\baselineskip}{10pt}

\bibliography{ReferenceSCI2}

\end{document}